\documentclass[twocolumn,amsmath,amssymb, aps,prb]{revtex4-2}

\usepackage{graphicx}
\usepackage{dcolumn}
\usepackage{bm}
\usepackage{subcaption}

\newcommand{\hethree}{$^3$He}
\newcommand{\hefour}{$^4$He}
\newcommand{\unit}[1]{\,\mathrm{#1}}
\begin{document}

\preprint{APS/123-QED}

\title{Superfluid \hethree-B Surface States in a  Confined Geometry\\ Probed by a Microelectromechanical Oscillator}

\author{W. G. Jiang}
\author{C. S. Barquist}%
\author{K. Gunther}
\author{Y. Lee}
\affiliation{Department of Physics, University of Florida}

\author{H. B. Chan}
\affiliation{Department of Physics, Hong Kong University of Science and Technology}


\date{\today}

\begin{abstract}
A microelectromechanical oscillator with a $0.73\unit{\mu m}$ gap structure is employed to probe the surface Andreev bound states in superfluid $^3$He-B. The surface specularity of the oscillator is increased by preplating it with 1.6 monolayers of $^4$He. In the linear regime, the temperature dependence of the damping coefficient is measured at various pressures, and the normalized energy gap is extracted. The damping coefficient increases after preplating at lower pressures, which is attributed to the decreased energy minigap of the surface bound states. The device is also driven into the nonlinear regime, where the temperature independent critical velocity at each pressure is measured. The critical velocity is observed to increase after preplating at all pressures, which might be related to the increased average energy gap. The observed behavior warrants a microscopic theory beyond a single parameter characterization of the surface.

\end{abstract}

\maketitle

\section{\label{sec:intro}Introduction}
Superfluid helium-3 is an unconventional superfluid in the p-wave spin triplet pairing state. It serves as a unique test-bed for theories of unconventional and topological superconductivity, garnering a wide range of interests.  In general, the order parameter of a non s-wave pairing is susceptible to defects and/or disorder. For example, the presence of a surface breaks the translational symmetry and consequently deforms the order parameter structure near the boundary of $\approx5-10$ coherence lengths from the surface. In particular, the bulk superfluid $^3$He-B phase with an isotropic energy gap becomes highly anisotropic with usuanl excitations near the boundary. These anisotropic states, emerging inside the bulk superfluid gap, are called the surface Andreev bound states (SABS), in which the quasiparticles are localized within the layer by Andreev scattering \cite{Nagato1998jltp}. The energy landscape of these midgap states are affected by the property of the surface wall, which is traditionally modeled by a single parameter, the average surface specularity $S$. $S$ represents the specular reflection probability averaged over the entire surface, where $S=0$ corresponds to a fully diffusive surface, and $S=1$ a fully specular one \cite{Nagato1996jltp, Nagato1998jltp, Murakawa2011jpsj}. \\

The B-phase is a rare material that represents a fully gapped 3D topological superfluid protected by symmetry.  Therefore, the topological properties in $^{3}$He-B are projected onto the SABS which received renewed attention.  Due to its unusual residual symmetry, many interesting phenomena have been predicted theoretically \cite{Wu2013,Mizushima2016jpsj}. Furthermore, the close proximity of two SABS, such as with $^3$He-B confined in a slab geometry, should exhibit a dependence of the order parameter on the size of the confinement, and is predicted to energetically favor new phases when the confinement reaches the size of the SABS \cite{Vorontsov2003, Vorontsov2007prl}.  \\

The SABS have been probed using various techniques such as nuclear magnetic resonance \cite{Levitin2013prl, Levitin2013science}, transverse acoustic impedance \cite{Aoki2005jltp, Aoki2005prl, Wada2008prb}, heat capacity \cite{Choi2006, Bunkov2020sr}, mechanical resonators, and many more. Mechanical resonators such as torsional pendulums \cite{Morley2002jltp, Parpia2003physb}, vibrating wires and grids \cite{Bradley2016nphys, Bradley2008jltp_grid}, and tuning forks \cite{Bradley2008jltp, Bradley2009, Clovecko2010jltp} immersed in bulk superfluid $^3$He-B have been shown to be successful detectors of the SABS. In the recent decades, the fast development of nanofabrication technology has given experimentalists the means to fabricate submicronic confinements with a stunning accuracy to reach a size comparable to the superfluid coherence length \cite{Levitin2013science, Rojas2015prb}. In particular, microelectromechanical system (MEMS) oscillators with large aspect ratios have been developed at the University of Florida to investigate the SABS \cite{Gonzalez2013, Gonzalez2016prb}. These devices are designed with a horizontal center plate suspended above the substrate, forming a gap structure beneath the plate. A more detailed description of the MEMS oscillators is given in Section~\ref{subsec:mems}.\\

In our previous work, we made extensive measurements using a MEMS oscillator with a $1.25\unit{\mu m}$ gap structure in pure $^{3}$He under diffusive boundary condition \cite{Zheng2014, Zheng2015jltp, Zheng2016prl, Zheng2017jltp}. By driving the MEMS oscillator in the linear regime, we observed a linear temperature dependent damping coefficient below $\sim 0.15T_c$, and successfully measured the suppression of the normalized energy gap in the SABS as a function of the scaled film thickness \cite{Zheng2017prl}.   By driving the MEMS oscillator further into the nonlinear regime, we observed a critical velocity much lower than the Landau critical velocity estimated for bulk liquid.  Unfortunately, during our follow-up experiment extending to the specular boundary condition, we lost the device.   In this work, we report the measurements using a MEMS oscillator with a $0.73\unit{\mu m}$ gap structure to investigate the effect of surface specularity.  We present and compare the results under two boundary conditions with pure $^{3}$He and with $1.6$ monolayers of $^4$He preplated which is the highest coverage we can attain with long term stability.  Beyond this coverage we observed the time dependent decrease of the coverage \cite{Jiang2023jpscp}.
\section{\label{sec:devices}Devices}

\subsection{\label{subsec:setup}Experimental setup}
\begin{figure}[b]
    \centering
    \includegraphics[width=0.3\textwidth]{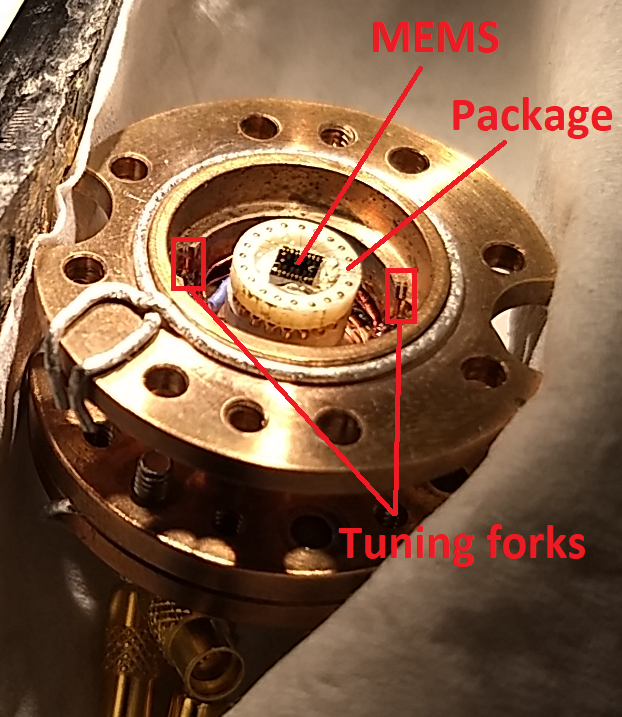}
    \captionsetup{justification=raggedright,singlelinecheck=false}
    \caption{The layout of the experimental cell. An NMR coil is mounted below the cell.}
    \label{fig:cell}
\end{figure}

Three parameters are controlled for this experiment: temperature and pressure of the superfluid $^3$He, and the surface specularity of the MEMS oscillator. The experimental cell is installed on a cryostat cooled by a combination of a dilution refrigerator and a copper demagnetization stage. The measurements are conducted during the natural warm-ups after the adiabatic demagnetization. The layout of the experimental cell is shown in Fig.~\ref{fig:cell}.
One MEMS chip and two tuning forks are installed in the experimental cell. There are 6 MEMS devices on the chip; one device is used for this experiment. The Pt NMR serves as the primary thermometer in superfluid $^3$He, and the tuning forks serve as secondary thermometers. The experimental cell was pressurized with liquid $^3$He to four different pressures of $3\,\mathrm{bar}$, $12\,\mathrm{bar}$, $25\,\mathrm{bar}$, and $29\,\mathrm{bar}$, where the MEMS oscillator and the tuning forks are submerged in the liquid.
The experiment is conducted in two different conditions: the non-preplate condition where the MEMS oscillator is in direct contact with pure $^3$He, and the preplated condition where the MEMS oscillator was coated with 1.6 monolayers of $^4$He \cite{Jiang2023jpscp}.

\subsection{\label{subsec:mems}MEMS oscillator}
\begin{figure}[b]
    \centering
    \begin{subfigure}[b]{0.45\textwidth}
        \includegraphics[width=1\linewidth]{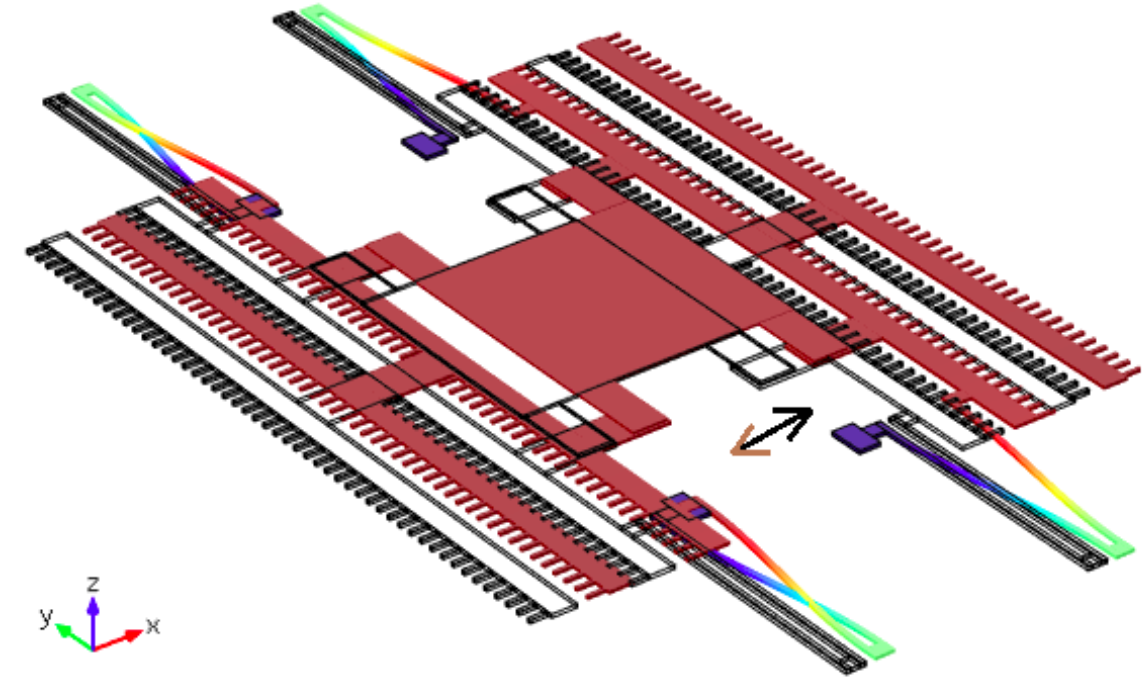}
        \subcaption{\label{fig:mems3d}}
    \end{subfigure}

    \begin{subfigure}[b]{0.4\textwidth}
        \includegraphics[width=1\linewidth]{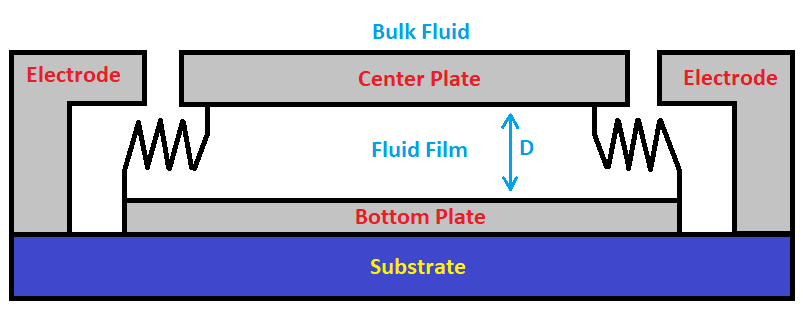}
        \subcaption{\label{fig:memsSide}}
    \end{subfigure}
    \captionsetup{justification=raggedright,singlelinecheck=false}
    \caption{\label{fig:mems}(a) A 3D diagram of the movable parts and the serpentine springs of the MEMS oscillator in action. The arrows indicate the direction of oscillation in the shear eigenmode. (b) A schematic sideview of the MEMS oscillator. The center plate is suspended above the substrate, forming a gap of thickness $D$ between them. When the oscillator is immersed in liquid \hethree, a bulk is formed above the plate, and a film below it.}
\end{figure}

The MEMS oscillator is fabricated at MEMSCAP with the PolyMUMPS process. A 3D diagram of the MEMS oscillator is shown in Fig.~\ref{fig:mems}(a).
The schematic drawing of the side view of the device structure is shown in Fig.~\ref{fig:memsSide}. The main component of the device is the movable part, which includes a center plate of $1.5
\times125\times125\,\mathrm{\mu m^3}$ and the attached comb-teeth electrodes, suspended above the substrate by four serpentine springs, thus forming a uniform gap between the plate and the substrate. The thickness of the gap is $0.73\,\mathrm{\mu m}$. Measurements below $35\,\mathrm{kHz}$ in vacuum at $4\,\mathrm{K}$ reveal four vibrational eigenmodes for the oscillator. In this work, we only focus on the shear mode, where the movable parts oscillate horizontally as indicated by the arrows in Fig.~\ref{fig:mems}(a). The resonance frequency of the shear mode in vacuum at $4\,\mathrm{K}$ is $23769\,\mathrm{Hz}$ with a quality factor of $\approx 6\times10^5$.

The measurement circuit for the MEMS oscillator is the ``push-pull'' capacitance bridge circuit \cite{Zheng2015jltp}. The comb-teeth structure on the oscillator allows its motion to be capacitively driven and detected given a DC bias. The driving force on the oscillator is
\begin{equation}
    F_{ex}=\tilde{\beta}V_DV_L,
    \label{eq:Fex}
\end{equation}
where $V_D=10\unit{V}$ is the constant DC bias used in this work, $V_L$ is the AC excitation, and
\begin{equation}
    \tilde{\beta}=\frac{N\varepsilon\varepsilon_0h}{d}
\end{equation}
is the transduction factor of the MEMS device. Here, $N=190$ is the number of parallel plate capacitors formed by the interdigitaed teeth on one side of the MEMS oscillator, $\varepsilon_0$ is the vacuum permittivity, $\varepsilon$ is the relative permittivity of the surrounding fluid, $h=5\,\mathrm{\mu m}$ is the thickness of the teeth, and $d=2\,\mathrm{\mu m}$ is the distance between the interdigitated teeth. The transduction factor is a geometrical factor with negligible temperature and pressure dependencies, and is calibrated in advance. The calibration in liquid $^3$He at $2.29\,\mathrm{bar}$, $1.66\,\mathrm{K}$ gives $\tilde{\beta}=1.98\times10^{-9}\,\mathrm{F/m}$.

\begin{figure}[b]
    \centering
    \includegraphics[width=\linewidth]{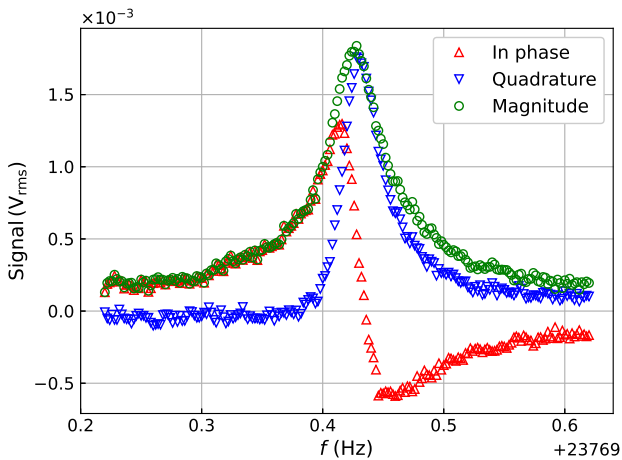}
    \captionsetup{justification=raggedright,singlelinecheck=false}
    \caption{An example frequency spectrum of the MEMS oscillator measured at $4\,\mathrm{K}$ in vacuum. The in-phase, quadrature, and magnitude data correspond to the real part, the imaginary part, and the modulus of $\mathcal{S}$.}
    \label{fig:memsPeak}
\end{figure}

An example resonance spectrum of the oscillator obtained by sweeping frequency in the linear regime is shown in Fig.~\ref{fig:memsPeak}. The example was measured at $4\,\mathrm{K}$ in vacuum. The spectrum is Lorentzian,
\begin{equation}
    \mathcal{S}(f)=\frac{A}{4\pi^2}\frac{f_o^2-f^2+i\tilde{\gamma}f}{(f_o^2-f^2)^2+(\tilde{\gamma}f)^2},
    \label{eq:Sf}
\end{equation}
where $A$ is the amplitude of the Lorentzian peak, $\tilde{\gamma}$ is the full-width-half-maximum (FWHM), and $f_o$ is the resonance frequency. The real and imaginary components correspond to the in-phase and the quadrature components in Fig.~\ref{fig:memsPeak},
while the modulus of $\mathcal{S}$ corresponds to the magnitude spectrum. Here, $\mathcal{S}$ is proportional to the displacement amplitude of the oscillatory motion of the movable parts. At $f=f_o$,
\begin{equation}
    x_{0r}=i\frac{\mathcal{S}_r}{\alpha\tilde{\beta}V_h},
\end{equation}
where $\mathcal{S}_r=\mathcal{S}(f_o)$, $x_{0r}$ is the amplitude of the displacement on resonance, $\alpha$ is a known amplification factor determined by the measurement circuit, and $V_h$ is the known voltage of a $150\,\mathrm{kHz}$ carrier wave signal. The oscillator can be driven into two different regimes: the linear and the nonlinear regime. In the linear regime, $x_{0r}$ is proportional to $F_{ex}$, and $\tilde{\gamma}$ is proportional to the damping coefficient of the fluid around the oscillator, while in the nonlinear regime, $x_{0r}$ is not proportional to $F_{ex}$. We will first present the data taken in the linear regime, and then in the nonlinear regime.


\section{\label{sec:data}Experiment and discussion}
Before presenting the data, it is helpful to first establish that the interaction between the superfluid \hethree{} quasiparticles and the MEMS oscillator is dominated by ballistic scatterings. The mean free path of the $^3$He quasiparticles near $T_c$ is $\lambda_c\approx4-40\,\mathrm{\mu m}$. Below $T_c$, $\lambda$ can be expressed as \cite{VollhardHe3}:
\begin{equation}
	\lambda(T)=\frac{(2\pi)^{1/2}}{3}\left(\frac{k_BT_c}{\Delta_o(0)}\right)^2\frac{1}{W_o}[v_F\tau_o(T_c)]e^{\Delta/k_BT},
	\label{eq:lambda}
\end{equation}
where $\Delta_o(0)$ is the \hethree{} zero temperature bulk energy gap, $v_F$ is the Fermi velocity, $W_o$ is a dimensionless parameter that depends only on the quasiparticle scattering amplitude in the normal phase, and $\tau_o$ is proportional to the lifetime of the quasiparticles. The bulk energy gap in the \hethree-B phase as a function of temperature is given by \cite{Halperin1990he3}
\begin{equation}
	\Delta_o(T)=\Delta_o(0)\tanh\left\{\frac{\pi k_BT_c}{\Delta_o(0)}\left[\left(\frac{T_c}{T}-1\right)\frac{2}{3}\frac{\Delta C}{C}\right]^{1/2}\right\}.
	\label{eq:DeltaT}
\end{equation}
$\Delta_o(T)/\Delta_o(0)$ is only weakly pressure dependent, and also weakly temperature dependent with $\Delta_o(T)>0.9\Delta_o(0)$ below $0.6T_c$. Therefore, in the temperature range reported in this work, $\Delta_o(T)$ can be reasonably treated as a constant, and $\lambda$  increases exponentially as $T$ decreases. At $T=0.5T_c$, $\lambda \approx 30\lambda_c$ which is on the order of $120\unit{\mu m}-1.2\unit{mm}$, and grows rapidly as the temperature decreases. The mean free path is thus comparable to or larger than the dimensions of the movable parts of MEMS oscillator, {\it i.e.} in the Knudsen regime, where the behavior of the fluid flow is dictated by the ballistic quasiparticle-wall scattering.

\subsection{\label{subsec:linear}The linear regime}
\begin{figure}[t]
    \centering
    \includegraphics[width=\linewidth]{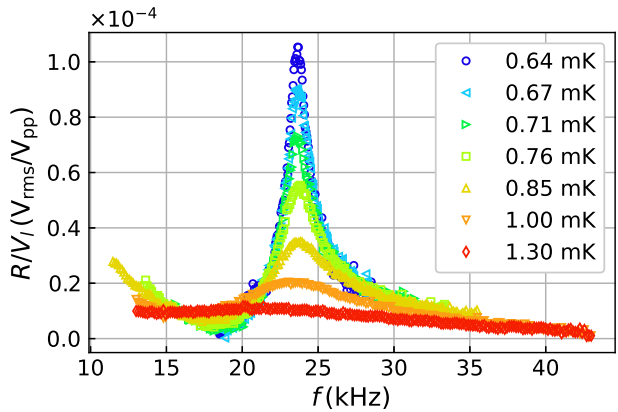}
    \captionsetup{justification=raggedright,singlelinecheck=false}
    \caption{A set of MEMS frequency sweeps at $28.8\unit{bar}$ with the non-preplated boundary condition. The signals are scaled by $V_L$. $\tilde{\gamma}$ increases as temperature increases.}
    \label{fig:fsweep_scaled}
\end{figure}

\begin{figure}[b]
    \centering
    \includegraphics[width=\linewidth]{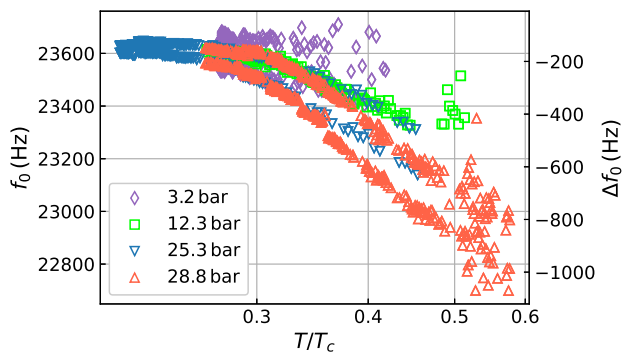}
    \captionsetup{justification=raggedright,singlelinecheck=false}
    \caption{The resonance frequency (left axis) and its shift from the vacuum value measured at $4\unit{K}$ (right axis) as a function of the reduced temperature in the log-linear scale.}
    \label{fig:f0_vs_T}
\end{figure}

The frequency spectrum of the MEMS oscillator in superfluid \hethree{} was first measured with the non-preplate boundary condition at various temperatures. A set of spectra taken at $28.8\unit{bar}$ with the non-preplated boundary condition are shown in Fig.~\ref{fig:fsweep_scaled}. Each spectrum is normalized by the driving force. 
The quality factor decreases from $\approx20$ at the lowest temperature to $\approx1$ before the superfluid A-B transition, when $\tilde{\gamma}$ can no longer be accurately identified. The values of $f_o$ and $\tilde{\gamma}$ are algorithmically extracted using Eq.~\ref{eq:Sf}. The measured $f_o$ as a function of the reduced temperature is shown in Fig.~\ref{fig:f0_vs_T}. The shift of $f_o$ from its vacuum value measured at $4\unit{K}$ is shown on the right axis.
$f_o$ is practically constant and pressure independent for $T<0.3T_c$, and the shift of $f_o$ at the highest temperature is quite small, less than $5\%$, which is consistent with the ballistic scattering picture, where the liquid is not viciously dragged by the oscillator as a continuum.

In the ballistic regime, the thermal damping experienced by the MEMS device is represented by \cite{Fisher1989}
\begin{equation}
	F_\mathrm{th}\,\propto\, T\exp(-\Delta/k_BT)[1-\exp(-\Lambda p_Fv/k_BT)],
	\label{eq:fisher_model}
\end{equation}
where $v$ is the velocity of the oscillator, and $\Lambda$ is a geometrical factor of the order of unity. The velocity of the device for frequency sweeps in the linear regime is $v\sim1\unit{mm/s}$. As an order of magnitude estimation for this experiment, $p_F\sim1\times10^{-24}\unit{kg\cdot m/s}$, and $T\sim1\unit{mK}$. Therefore, $p_Fv\sim1\times10^{-27}\unit{J}$, and $k_BT\sim1.38\times10^{-26}\unit{J}\gg p_F v$, which implies that the MEMS oscillator is in the low velocity limit. In this limit, Eq.~\ref{eq:fisher_model} can be approximated to the first order expansion of $v$ as
\begin{equation}
	F_\mathrm{th}\,\propto\,\frac{\Lambda p_F}{k_B}\exp(-\Delta/k_BT)\cdot v.
	\label{eq:fisher_v_expand}
\end{equation}
This gives a damping coefficient that exponentially decreases with temperature, which has been observed with vibrating wires, and tuning forks \cite{Guenault1986, Bradley2009}.

\begin{figure}[t]
    \centering
    \includegraphics[width=\linewidth]{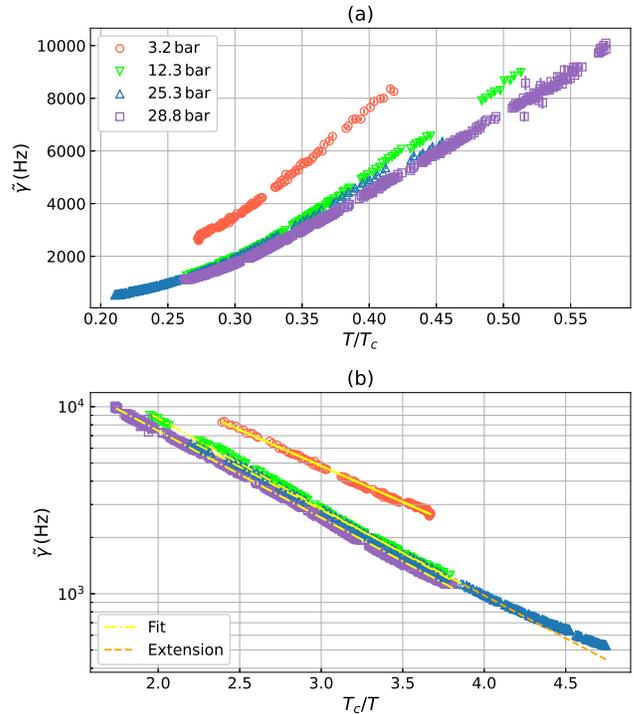}
    \captionsetup{justification=raggedright,singlelinecheck=false}
    \caption{(a) $\tilde{\gamma}$ as a function of $T/T_c$. (b) $\tilde{\gamma}$ as a function of $T_c/T$ in the linear-log scale. The data are fitted in the temperature range of $T/T_c>0.26$ ($T_c/T<3.8$) using Eq.~\ref{eq:fisher_normalize}. The orange dashed line is the extension of the yellow dash-dot line down to the lowest temperature at $25.3\unit{bar}$. The curve deviates from the exponential dependence above $T_c/T\approx4.2$.}
    \label{fig:ana_dvsT}
\end{figure}

\begin{figure}[b]
    \centering
    \includegraphics[width=0.8\linewidth]{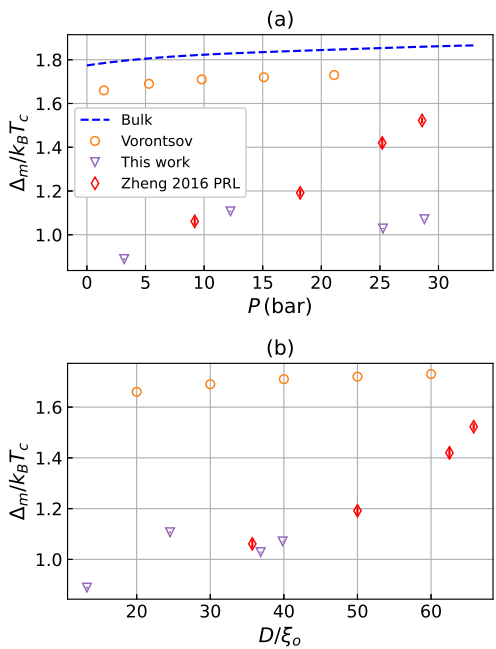}
    \captionsetup{justification=raggedright,singlelinecheck=false}
    \caption{\label{fig:DeltaKT_vs_DXi}The measured $\Delta_m/k_BT_c$ as a function of (a) pressure, and (b) the scaled film thickness $D/\xi_o$. The purple triangles are measured in this work using a MEMS device that defines a $0.73\unit{\mu m}$ film beneath the center plate. The red diamonds are measured in a previous work \cite{Zheng2016prl} using a similar MEMS device that defines a $1.25\unit{\mu m}$ film. The blue dashed line is the zero temperature bulk energy gap calculated using the weak coupling plus model. The orange circles are the average energy gap in a $1.25\unit{\mu m}$ film trapped between two diffusive walls calculated by Vorontsov \cite{Zheng2016prl}.}
\end{figure}

The temperature dependence of $\tilde{\gamma}$ as a function of $T/T_c$ is shown in Fig.~\ref{fig:ana_dvsT}(a), and Fig.~\ref{fig:ana_dvsT}(b) is the Arrhenius plot of the same data.
For all pressures and $T_c/T<3.8$, $\tilde{\gamma}$ takes a linear shape in Fig.~\ref{fig:ana_dvsT}(b), as expected from Eq.~\ref{eq:fisher_v_expand}. The thermal damping coefficient, $\tilde{\gamma}_\mathrm{th}$ can be fitted to a form commensurate with Eq.~\ref{eq:fisher_v_expand}:
\begin{equation}
	\tilde\gamma_\mathrm{th}=\mathcal{A}\exp(-\frac{\mathcal{C}}{T/T_c}).
	\label{eq:fisher_normalize}
\end{equation}
with $\mathcal{A}$ and $\mathcal{C}$ as fitting parameters. Specifically, $\mathcal{C}=\Delta_m/k_BT_c$, where $\Delta_m$ is the measured energy gap. The fit is done for temperatures $0.26 < T/T_c < 3.8$ and plotted as yellow dash-dot lines in Fig.~\ref{fig:ana_dvsT}(b). The pressure dependence of the measured $\Delta_m/k_BT_c$ is displayed in Fig.~\ref{fig:DeltaKT_vs_DXi}(a) along with the previous measurements with a similar MEMS oscillator with a $1.25\unit{\mu m}$ gap structure \cite{Zheng2016prl}. The purple triangles are the measurements in this work, and the red diamonds are the measurements in the previous work. Two theoretical calculations are also plotted for comparison. The blue dashed line is the zero temperature bulk energy gap calculated using the weak coupling model \cite{Halperin1990he3}. The orange circles are the average energy gap in a $1.25\unit{\mu m}$ film trapped between two diffusive walls calculated by Vorontsov \cite{Zheng2016prl}. The experimental values of $\Delta/k_BT_c$ are lower than the theoretical values. However, they all increases with pressure. The physically meaningful parameter is the scaled film thickness, $D/\xi_o$ rather than the pressure to accommodate the geometric difference between the devices.  The two sets of measured $\Delta_m/k_BT_c$ follow the same dependence despite the big difference in the thickness. 
Nagato {\it et al.} calculated the self-consistent order parameter and surface density of states at $T=0.2T_c$ at various surface boundary conditions \cite{Nagato1998jltp}. Nagato's calculation shows that the order parameter is severely suppressed on a surface, and recovers to its bulk value over a distance of $\sim 5\xi_o$. As $D$ decreases, the region with suppressed $\Delta$ occupies a larger proportion of the film, resulting in a smaller average energy gap.

\begin{figure}[t]
    \centering
    \includegraphics[width=\linewidth]{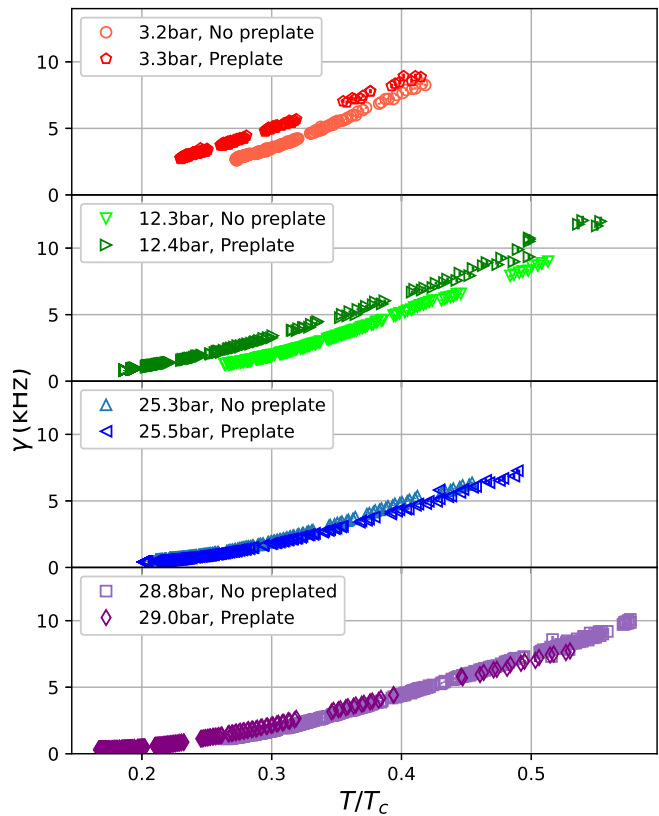}
    \captionsetup{justification=raggedright,singlelinecheck=false}
    \caption{\label{fig:gamma_preplate} $\tilde{\gamma}$ measured at various pressures with both boundary conditions.}
\end{figure}

\begin{figure}[t]
    \centering
    \includegraphics[width=0.8\linewidth]{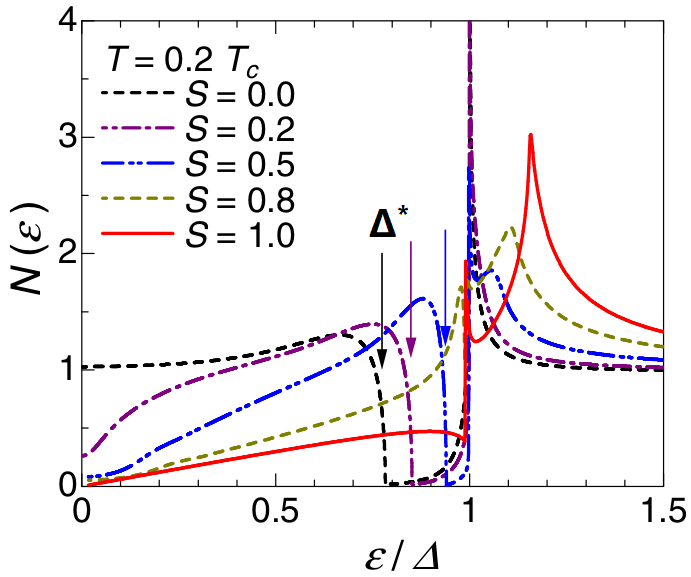}
    \captionsetup{justification=raggedright,singlelinecheck=false}
    \caption{\label{fig:sdos} The Theoretical calculation of the SDoS as a function of the quasiparticle energy $\varepsilon$ for \hethree-B with various specularities, $S$, at $T=0.2T_c$ . Here, $S=0$ represents a fully diffusive surface, while $S=1$ is a specular surface. The SDoS is normalized by the normal state value, and $\varepsilon$ is normalized by the superfluid energy gap. The arrows point to the band edge energy $\Delta^*$ . Reproduced from Ref.~\cite{Murakawa2011jpsj}.}
\end{figure}

A key result of Nagato's calculation is that the suppression of $\Delta$ heavily depends on the surface boundary condition: $\Delta$ increases as the surface specularity increases. Therefore, we measured $\tilde{\gamma}$ at the same pressures with the MEMS oscillator preplated by 1.6 monolayers of \hefour.  The \hefour{} coverage was determined directly by measuring the resonance frequency shift after the preplating \cite{Jiang2023jpscp}.  The measured $\tilde{\gamma}$ with both boundary conditions are shown in Fig.~\ref{fig:gamma_preplate}.
At $3\unit{bar}$ and $12\unit{bar}$, $\tilde{\gamma}$ increased after preplating the MEMS oscillator, while the same increase was not observed at $25\unit{bar}$ and $29\unit{bar}$. The diminishing augmentation of surface specularity by a fixed \hefour{} coverage at higher pressures has been universally observed \cite{Steel1990, Tholen1991, Einzel1997, Matsubara1999jltp}. Therefore, it is not unexpected that the effect of surface specularity on $\tilde{\gamma}$ only manifests at the lower pressures. An enhanced surface specularity means that a larger proportion of quasiparticles undergo specular reflection on the MEMS oscillator, which lowers the average momentum transfer from each scattering. On the other hand, Nagato {\it et al.} calculated the surface density of states (SDoS) as a function of the quasiparticle energy $\varepsilon$ for \hethree-B with various specularities $S$ \cite{Murakawa2011jpsj}, shown in Fig.~\ref{fig:sdos}.
The SDoS is normalized by its normal state value, and $\varepsilon$ is normalized by the superfluid energy gap. Due to the distortion of the order parameter near the surface, midgap states that are prohibited in the bulk superfluid are now opened up. The arrows indicate the band edge gap $\Delta^*$. Because the suppression of $\Delta$ near the surface is anisotropic, where the perpendicular component $\Delta_\perp$ is completely suppressed irrespective of the surface specularity, the quasiparticle moves with the normal fluid Fermi velocity $v_\perp\approx v_F\approx50\unit{m/s}$. In the direction parallel to the surface, the energy gap $\Delta_\parallel$ limits the quasiparticle velocity to the Landau velocity $v_\parallel\approx v_L\sim 50\unit{mm/s}\ll v_F$. The thickness of the SABS can be estimated to be $\sim 5\xi_o\approx100\unit{nm}$. This means that the SABS quasiparticle undergoes $\sim10^4$ collisions during one cycle of the MEMS oscillation. This multi-collision process allows the MEMS oscillator to pump energy into the quasiparticles up to $\Delta^*$. When $S$ increases, the energy gap $\Delta-\Delta^*$ gradually closes, which opens a channel to promote the SABS quasiparticles above $\Delta$. Therefore, as $S$ is increased, a larger channel for promotion is opened up, resulting in an increased number of quasiparticles escaping the SABS which dissipates momentum, causing $\tilde{\gamma}$ to increase as observed.

\subsection{\label{subsec:nonlinear}The nonlinear regime}
\begin{figure}[t]
    \centering
    \includegraphics[width=0.9\linewidth]{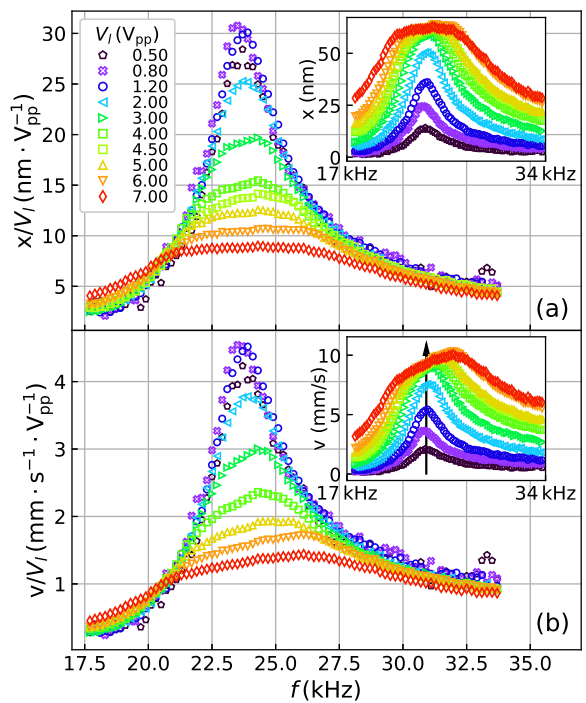}
    \captionsetup{justification=raggedright,singlelinecheck=false}
    \caption{\label{fig:nonlinear_sweeps} The frequency spectra of the shear mode of the MEMS oscillator at $0.75\unit{mK}$, $28.9\unit{bar}$ when the driving excitation, $V_L$, is increased into the nonlinear regime. (a) The spectra of the displacement amplitude of the movable parts normalized by the driving excitation. Inset: the displacement amplitude of the movable parts. (b) The spectra of the velocity normalized by the driving excitation. Inset: the spectra of the velocity of the movable parts. The black arrow indicates the direction of a excitation sweep.}
\end{figure}

When a sufficiently strong excitation is applied to the MEMS device, the oscillator enters the nonlinear regime where the displacement of the movable parts in the shear mode is no longer proportional to the excitation. A series of frequency spectra of the displacement amplitude measured with various excitation voltages, $V_L$ at $0.75\unit{mK}$, $28.9\unit{bar}$ are shown in the inset of Fig.~\ref{fig:nonlinear_sweeps}(a).
The spectra are no longer Lorentzian, and the displacement amplitude saturates at $\approx60\unit{nm}$. The displacement amplitude normalized by the driving excitation is shown in Fig.~\ref{fig:nonlinear_sweeps}(a). The peak structures are truncated, while the low amplitude signals (the left and right tails of the spectra) overlap, indicating that the low amplitude motion of the MEMS oscillator is still in the linear regime. The spectra of the velocity of the movable parts, and the velocity normalized by excitation are shown in Fig.~\ref{fig:nonlinear_sweeps}(b). The velocity spectra show similar features as that of the displacement.

\begin{figure}[t]
    \centering
    \includegraphics[width=0.9\linewidth]{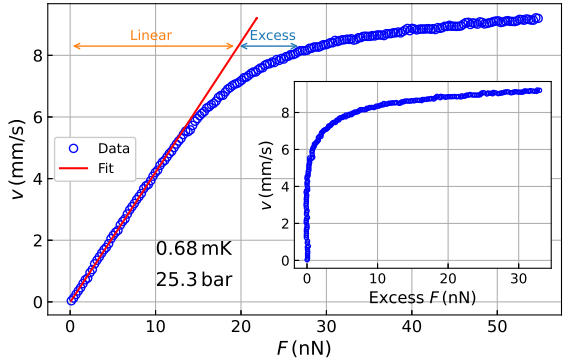}
    \captionsetup{justification=raggedright,singlelinecheck=false}
    \caption{\label{fig:vsweep_decompose_25bar} An excitation sweep taken at $0.68\unit{mK}$, $25.3\unit{bar}$ with the driving force being ramped up. The red solid line is linear fit to the low velocity part of the curve. The force can be decomposed into two components: the linear damping that was measured in the linear regime, and the excess damping force. The inset plot shows the relationship between the velocity and the excess damping force.}
\end{figure}

\begin{figure}[t]
    \centering
    \includegraphics[width=0.9\linewidth]{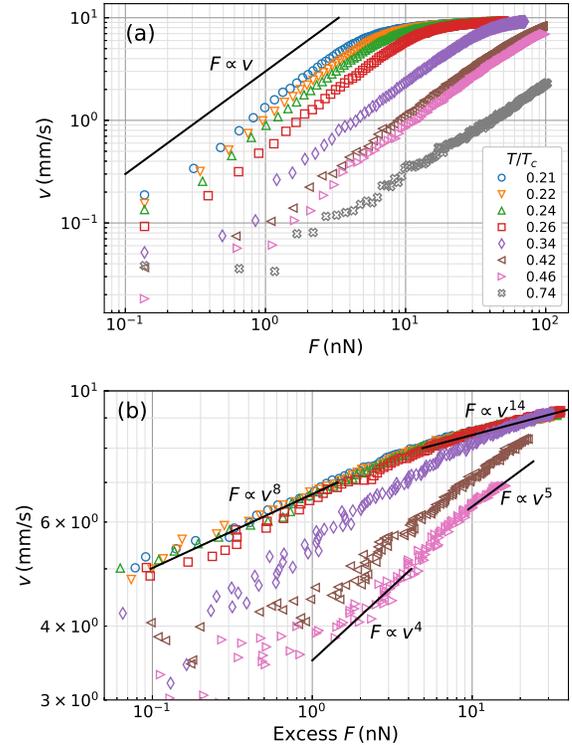}
    \captionsetup{justification=raggedright,singlelinecheck=false}
    \caption{\label{fig:vsweep_varyT_25bar} The $v$ dependence of the damping force on the MEMS oscillator at $25.3\unit{bar}$ measured with the driving force being ramped up. (a) A series of excitation sweeps at various temperatures. $v$ saturates at a temperature independent critical velocity of $v_c\approx9.2\unit{mm/s}$. As $T/T_c$ increases, the slope of the linear part decreases, indicating an increasing $\tilde{\gamma}$. (b) The relationship between $v$ and the excess force. The $T/T_c=0.74$ curve is purely linear, and no excess force, hence not included here.}
\end{figure}

In order to investigate the velocity dependence of the damping force, the MEMS oscillator is driven on resonance while the excitation force is varied, as indicated by the black arrow in the inset of Fig.~\ref{fig:nonlinear_sweeps}(b). A typical excitation sweep taken at $0.68\unit{mK}$, $25.3\unit{bar}$ is shown in Fig.~\ref{fig:vsweep_decompose_25bar} where the driving force was ramped up.
The $v-F$ relation curve has a linear low-velocity part. A linear fit is performed on the linear part, shown as the solid red line in the figure. The red line decomposes the damping force into two components: the linear damping where $F\propto v$, and the excess damping force with a higher order $v$ dependence. The slope of the linear part of the curve is the reciprocal damping coefficient, $\tilde{\gamma}^{-1}$, in the linear regime. The excess damping force is obtained by subtracting the linear damping from the total damping, shown as the inset in Fig.~\ref{fig:vsweep_decompose_25bar}. The $v$ dependence of the damping force at various temperatures, $25.3\unit{bar}$ are shown in the log-log scale in Fig.~\ref{fig:vsweep_varyT_25bar}(a). Again, the curves were swept with the driving force being ramped up.
The driving force is limited by the electric breakdown voltage on the comb-teeth capacitors. The low velocity linear dependence on $F$ is evident, as illustrated by the $F\propto v$ guideline. As $T$ increases, $\tilde{\gamma}$ increases, shown by the decreasing slope of the linear part. At high temperatures, the damping is purely linear in the measurement range, while at low temperatures, $v$ saturates to a temperature independent critical velocity of $v_c\approx9.2\unit{mm/s}$. The $v$ dependence of the excess force is plotted in Fig.~\ref{fig:vsweep_varyT_25bar}(b). The low velocity noise for $v<3\unit{mm/s}$ are excluded for clarity. The excess force has a high power $v$ dependence: $F\propto v^j$, where $j>1$. While $j$ increases with $F$ at a given $T$, $j$ increases with $T$ at a given $v$. The power law dependence for the low and high velocity parts of the excess force are estimated, denoted by the guidelines in Fig.~\ref{fig:vsweep_varyT_25bar}(b). $j$ varies from 4 to 14.

\begin{figure}[t]
    \centering
    \includegraphics[width=\linewidth]{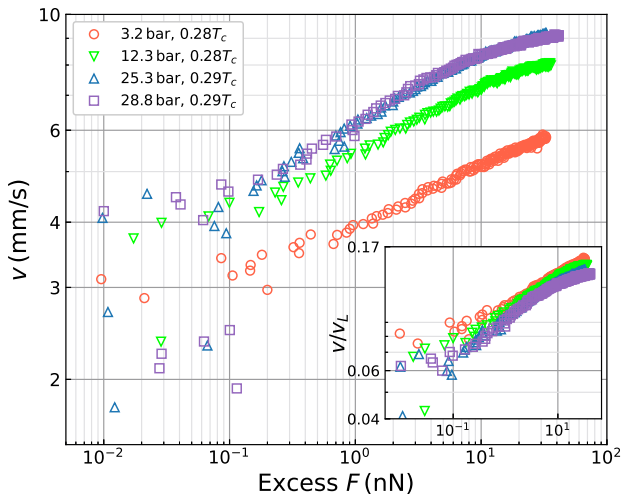}
    \captionsetup{justification=raggedright,singlelinecheck=false}
    \caption{\label{fig:vsweep_028Tc} The relationship between $v$ and the excess force at $T=0.28T_c-0.29T_c$ and various pressures are plotted in the log-log scale. (Inset) The relationship between $v/v_L$ and the excess force. The curves collapse after normalizing the velocity. The x-axis covers the same range of excess force as that of the main graph.}
\end{figure}

\begin{figure}[b]
    \centering
    \includegraphics[width=0.8\linewidth]{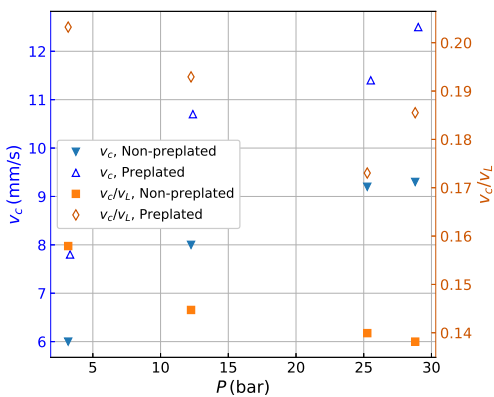}
    \captionsetup{justification=raggedright,singlelinecheck=false}
    \caption{\label{fig:vcvL_P} The pressure dependence of $v_c$ (left y-axis) and $v_c/v_L$ (right y-axis) with both boundary conditions. The solid markers represent data taken in pure \hethree, and the hollow markers with the preplated boundary condition. $v_c$ increases with $P$ for both boundary conditions, while $v_c/v_L$ are less pressure dependent. Increasing the surface specularity increases $v_c$ and $v_c/v_L$.}
\end{figure}

The relationship between $v$ and the excess force at $T=0.28T_c-0.29T_c$ and various pressures are plotted in the log-log scale in Fig.~\ref{fig:vsweep_028Tc}.
Given the same $v$ at the same $T/T_c$, the excess force is larger at higher pressures. The inset in Fig.~\ref{fig:vsweep_028Tc} depicts the relationship between the excess force and the velocity normalized by the Landau critical velocity $v_L$. The four curves approximately collapse into the same curve after normalization. This universal scaling suggests that the excess force originates from quasiparticle pair breaking above the energy gap. The pressure dependence of $v_c$ and $v_c/v_L$ are shown in Fig.~\ref{fig:vcvL_P}.
$v_c$ has a strong pressure dependence; it increases by $\approx50\%$ from 3 to $29\unit{bar}$. In contrast, $v_c/v_L$ has a much weaker pressure dependence; it decreases from $\approx0.16$ to $\approx0.14$ as the pressure increases. The observed $v_c/v_L$ is much lower than the observed value of $\approx0.25$ by Castelijin {\it et al.} with a vibrating wire \cite{Castelijns1986prl}. The discrepancy could not be explained by the geometrical difference between the two oscillators. Lambert pointed out that the measured critical velocity is suppressed by a factor of $1+\alpha_g$ where $\alpha_g$ is a geometrical factor \cite{Lambert1992}. For a tuning fork, $\alpha_g=2$, while for a thin plate in its shear motion, $\alpha_g\approx1$. This indicates that one should observe a higher $v_c/v_L$ using a MEMS oscillator. The previous experiment by Zheng {\it et al.} observed $v_c/v_L\approx0.08$ using the MEMS oscillator that has a $1.25\unit{\mu m}$ gap structure \cite{Zheng2017prl}. The unusually low critical velocity was speculated to stem from the microscopic structures of the SABS. In the SABS, the existence of the midgap states narrows the energy gap to $\Delta-\Delta^*$ as shown in Fig.~\ref{fig:sdos}. For a diffusive boundary, $\Delta-\Delta^*\approx0.2\Delta$, hence $v_c\approx0.2v_L/(1+1)=0.1v_L$ for a thin plate. This ratio is of the same order of magnitude as the one measured in this work.

\begin{figure*}
\includegraphics[width=\linewidth]{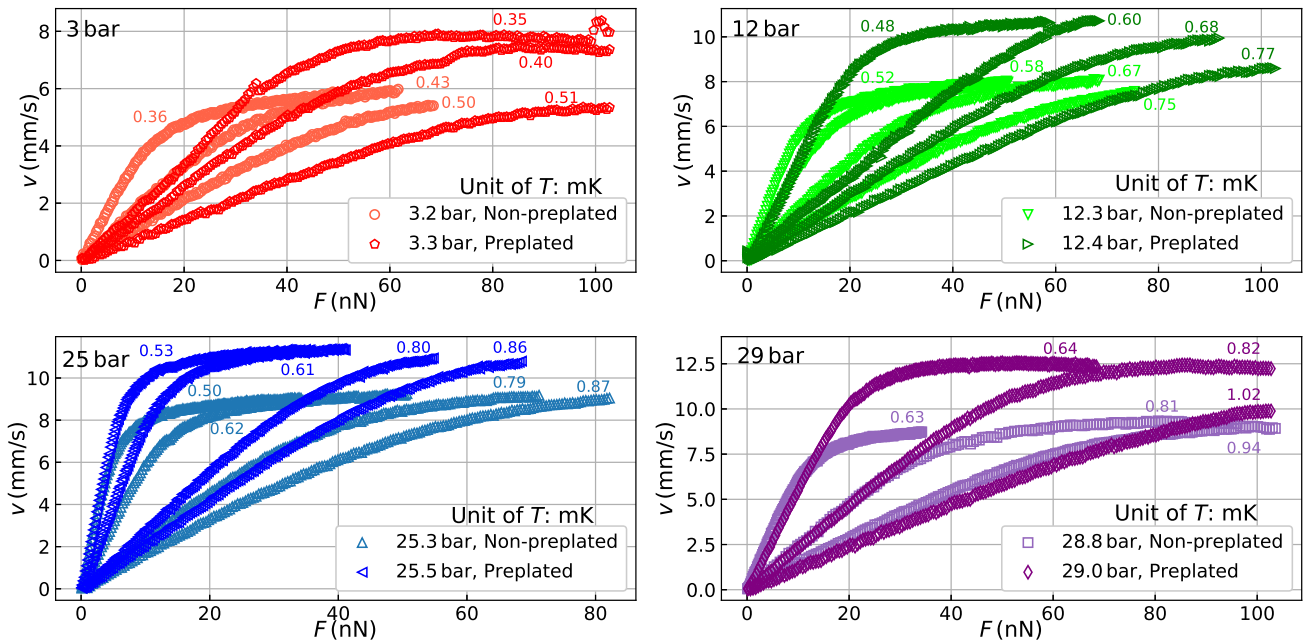}
\captionsetup{justification=raggedright,singlelinecheck=false}
\caption{\label{fig:vsweep_bounds}The excitation sweeps at various pressures for both boundary conditions. All sweeps were done with the excitation force being ramped up. The light red, light green, light blue, and light purple symbols correspond to the non-preplated boundaries; the red, green, blue, and purple symbols correspond to the preplated boundaries. The temperatures for each curve is labeled next to it. At the same pressure, curves at similar temperatures but different boundary conditions are plotted for direct comparison. The curves with the preplated boundary condition have smaller slopes for the linear part at $3$ and $12\unit{bar}$, and saturate at larger critical velocity $v_c$ at all pressures.}
\end{figure*}

The excitation sweeps at various pressures for both boundary conditions are depicted in Fig.~\ref{fig:vsweep_bounds}. At each pressure, the curves are chosen to include pairs of similar temperatures, one with each boundary condition. For example, at $3\unit{bar}$, a pair of curves are plotted where one is taken at $0.36\unit{mK}$ with the non-preplated boundary condition, and the other at $0.35\unit{mK}$ with $^{4}$He preplated. At $3$ and $12\unit{bar}$, the slope of the linear part is smaller with preplating; at $25$ and $29\unit{bar}$, no such discrepancy exist. This corroborates with the observation in the linear regime that $\tilde{\gamma}$ is only sensitive to preplating at the two lower pressures. However, $v_c$ is consistently larger at all pressures for the preplated boundary condition. The values of $v_c$ and $v_c/v_L$ are also plotted in Fig.~\ref{fig:vcvL_P}. Preplating increases $v_c$ to about $0.19v_L$. When the specularity increases, so does $\Delta$ \cite{Nagato1998jltp}, and larger $v_c$ is necessary to overcome the larger $\Delta$.

Another consequence of preplating on $\tilde{\gamma}$ in the linear regime and $v_c$ in the nonlinear regime at $25\unit{bar}$ and $29\unit{bar}$ may result from the different behaviors of the energy gaps $\Delta_\perp$ and $\Delta_\parallel$. The suppression of $\Delta_\perp$ near a surface is topologically protected, and insensitive to $S$, while the distortion of $\Delta_\parallel$ is sensitive to $S$: $\Delta_\parallel$ increases with $S$ \cite{Nagato1998jltp}. If $\tilde{\gamma}$ and $v_c$ are more susceptible to different components of $\Delta$, then they will have different sensitivity to the change of $S$. On the other hand, one could also argue that a single parameter $S$ is insufficient at describing a complex 3D wall. To define $S$, one requires that the Fermi wavelength of the superfluid be much longer than the height of the irregularities on the surface wall \cite{Nagato1996jltp}. In this work, the Fermi wavelength can be estimated to be
\begin{equation}
    \lambda_F=\frac{\hbar}{m^*v_F}\approx\frac{\hbar}{5m_3\times 40\unit{m/s}}\approx1\unit{\AA},
\end{equation}
where $m^*$ and $m_3$ are the masses of the superfluid $^3$He quasiparticle and the bare $^3$He atom respectively, and $v_F$ is the Fermi velocity. The surface irregularities on the movable parts of the MEMS oscillator were mapped by AFM; they have heights of $\sim10\unit{nm}$, and lateral sizes of $\sim100\unit{nm}$ \cite{Gonzalez2013}. Therefore, the irregularities are macroscopic compared to the Fermi wavelength, which could increase the proportion of quasiparticles being retroreflected, and means that the surface can no longer be characterized by $S$ alone.

\section{Conclusion}\label{sec:conclusion}
We extended a previous experiment to probe the SABS in superfluid $^3$He-B by using a MEMS oscillator that has a $0.73\unit{\mu m}$ gap structure compared to the previous $1.25\unit{\mu m}$. The surface specularity is controlled with the $^4$He preplating technique. In the linear regime of the oscillator, the temperature dependence of the damping coefficient was measured, and the normalized energy gap was extracted. The measured $\Delta_m/k_BT_c$ is consistent with the previous measurement. The damping coefficient increases when the specularity is increased. This is explained by the closing energy minigap, which establishes a more efficient channel for quasiparticle promotion from the midgap states in the SABS into the bulk superfluid. In the nonlinear regime, a temperature independent critical velocity is observed. The critical velocity at various pressures scales with the Landau critical velocity. It increases at all pressures after preplating, which is attributed to the tendency of the energy gap to increase with the specularity. It is noticed that the damping coefficient is only affected by the surface specularity at lower pressures, while the critical velocity is affacted at all pressures. We propose that $\Delta$ and $S$, both are average values, are insufficient to fully describe a system where the surface irregularities are larger than the atomic scale, and that more detailed theories are needed.

\begin{acknowledgments}
This work is supported by DMR-1708818 through the National Science Foundation.
\end{acknowledgments}


\bibliography{JWGBib}

\end{document}